# The primitive ontology of quantum physics: guidelines for an assessment of the proposals

Michael Esfeld

University of Lausanne, Department of Philosophy

Michael-Andreas.Esfeld@unil.ch

**Abstract**

The paper seeks to make progress from stating primitive ontology theories of quantum physics – notably Bohmian mechanics, the GRW matter density theory and the GRW flash theory – to assessing these theories. Four criteria are set out: (a) internal coherence; (b) empirical adequacy; (c) relationship to other theories; (d) explanatory value. The paper argues that the stock objections against these theories do not withstand scrutiny. Its focus then is on their explanatory value: they pursue different strategies to ground the textbook formalism of quantum mechanics, and they develop different explanations of quantum non-locality. In conclusion, it is argued that Bohmian mechanics offers a better prospect for making quantum non-locality intelligible than the GRW matter density theory and the GRW flash theory.

## 1. The primitive ontology theories of quantum mechanics

Suppose that there is matter distributed in three-dimensional space or four-dimensional space-time and that the task of physics is to develop an account of matter and its temporal development (plus an account of space and time themselves). If one endorses this supposition in the context of quantum physics, one is committed to what is known as a primitive ontology.[1] The ontology is primitive in the sense that it cannot be inferred from the formalism of textbook quantum mechanics (QM), but has to be put in as the referent of that formalism. The motivation for doing so is to obtain an ontology that can account for the existence of measurement outcomes – and, in general, the existence of the macrophysical objects with which we are familiar before doing science. Hence, what is introduced as the primitive ontology has to be such that it can constitute measurement outcomes and localized macrophysical objects in general. That is why the primitive ontology consists in one actual distribution of matter in space at any time (no superpositions), and the elements of the primitive ontology are localized in space-time, being "local beables" in the sense of Bell (2004, ch. 7), that is, something that has a precise localization in space at a given time.[2]

---

[1] This term goes back to Dürr, Goldstein and Zanghì (2013, ch. 2, see end of section 2.2, paper originally published 1992).

[2] The Everett-style primitive ontology set out in Allori et al. (2011) is an exception, but it is then doubtful what the motivation is to advance a primitive ontology instead of simply going for an ontology that admits only the quantum state of the universe (i.e. the universal wave-function).

There are three elaborate primitive ontology theories of QM discussed in today's literature. The de Broglie-Bohm theory, going back to de Broglie (1928) and Bohm (1952) and known today as Bohmian mechanics (BM) (see Dürr, Goldstein and Zanghì 2013), is the oldest of them. BM endorses particles as the primitive ontology, adding the position of the particles as additional variable to the formalism of textbook QM. Consequently, there is at any time one actual configuration of particles in three-dimensional space, and the particles move on continuous trajectories. BM therefore needs two laws: the guiding equation fixing the temporal development of the position of the particles, and the Schrödinger equation determining the temporal development of the wave-function. These two laws are linked in the following manner: the role of the wave-function, developing according to the Schrödinger equation, is to determine the velocity of each particle at any time *t* given the position of all the particles at *t*.

Furthermore, there are two primitive ontology theories of QM within the framework of the dynamics going back to Ghirardi, Rimini, Weber (GRW) (1986), which seeks to include the textbooks' postulate of the collapse of the wave-function upon measurement into a modified Schrödinger equation. Ghirardi, Grassi and Benatti (1995) set out an ontology of a continuous matter density distribution in physical space (GRWm): the wave-function in configuration space and its temporal development as described by the GRW equation represent at any time the density of matter in physical space, with matter being conceived as stuff (gunk) that is distributed throughout space. The spontaneous localization of the wave-function in configuration space (its "collapse") represents a spontaneous contraction of the matter density in physical space, thus accounting for measurement outcomes and well localized macrophysical objects in general (see also Monton 2004).

The other theory goes back to Bell (2004, ch. 22, originally published 1987): whenever there is a spontaneous localization of the wave-function in configuration space, that development of the wave-function in configuration space represents an event occurring at a point in physical space. These point-events are known as flashes; the term "flash", however, is not Bell's, but was introduced by Tumulka (2006, p. 826). According to the GRW flash theory (GRWf), the flashes are all there is in space-time. Consequently, the temporal development of the wave-function in configuration space does not represent the distribution of matter in physical space. It represents the objective probabilities for the occurrence of further flashes, given an initial configuration of flashes. Hence, there is no continuous distribution of matter in physical space, namely no trajectories or worldlines of particles, and no field – such as a matter density field – either. There only is a sparse distribution of single events in space-time. Although GRWf and GRWm are rival proposals for an ontology of the same formalism (the GRW quantum theory), there also is a difference between them on the level of the formalism: if one endorses the GRWm ontology, it is reasonable to pursue a formalism of a *continuous* spontaneous localization of the wave-function (as done in Ghirardi, Pearle and Rimini 1990), whereas if one subscribes to the GRWf ontology, there is no point in doing so.

There hence are three well worked out proposals for a primitive ontology of QM: particles as in BM, a matter density field as in GRWm, and flash-events as in GRWf. Each of these proposals comes with a different law: a guiding equation for the motion of particles in BM, an equation of a continuous spontaneous localization of the wave-function in GRWm representing the development of the matter density, and an equation of a jump-like localization of the wave-function in GRWf representing the occurrence of further flashes. That difference notwithstanding, the structure of these theories is the same (see Allori et al. 2008): there is a distribution of matter in physical space (the primitive ontology) and a law for its temporal development. On this basis, one can conceive many more primitive ontology theories of QM, combining elements from the three mentioned ones (see Allori et al. 2008 and 2014). Nonetheless, this paper is limited to considering BM, GRWm and GRWf. They are the three theories that are developed in detail, and they cover the main conceptions of matter: particles or atoms (BM), a field in the sense of stuff or gunk (GRWm), single events (GRWf).

Given the structure of these theories consisting in a proposal for a primitive ontology and a law for its temporal development, it would not make sense to regard the wave-function as a physical entity in addition to the primitive ontology: the wave-function could not be a field in physical space, but only a field in the high-dimensional mathematical space each point of which represents a possible configuration of matter in physical space. However, this relationship of representation could not be turned into a causal relation: it would be entirely mysterious how an entity existing in configuration space could exert a physical influence on the motion of entities existing in physical space. This is the main motivation for regarding the universal wave-function (that is, the wave-function of the whole configuration of matter in the universe) as nomological, namely as being part of the law of the development of the primitive ontology, by contrast to being a physical entity on a par with the primitive ontology (see Dürr, Goldstein and Zanghì 2013, chs. 11.5 and 12, in the context of BM).

Nonetheless, it is also reasonable to take the wave-function to encode a physical fact over and above the primitive ontology, given in particular that the primitive ontology is not sufficient to determine the wave-function: there can be two identical universal configurations of particles, matter density or flashes and yet different wave-functions of these configurations, leading to different temporal developments of them. Adopting Bell's terminology, one could simply say that the wave-function represents a "non-local beable" that exists over and above the "local beables" making up the primitive ontology. But such a statement alone would not be informative: it is clear what a "local beable" is, because this can be spelled out in terms of particles, events (flashes), field values, etc. existing at points in physical space. If one admits over and above such "local beables" a "non-local beable" represented by the universal wave-function in the ontology, one also has to spell out what such a "non-local beable" is.

One possibility to fill this lacuna is provided by the view that the wave-function represents a property of matter (see Monton 2006). More precisely, the role that the wave-function plays in the law for the temporal development of the distribution of matter captures a property of the whole configuration of matter, namely a disposition or propensity that manifests itself in a certain development of that configuration and that thereby grounds the law (the guiding equation in BM, the GRW equation) in the sense that the law supervenes on this property.[3] Another possibility is to adopt primitivism with respect to that law (see Maudlin 2007), that is, take the view that over and above a configuration of matter, there is in each world the fact instantiated that a certain law holds in this world. I shall make use of the property view of the wave-function in the following, but nothing hangs on this view for present purposes. When this view is employed, one can also put in primitivism instead.[4]

---

[3] See Belot (2012, pp. 77-80) and Esfeld et al. (2013, sections 4-5) for BM as well as Dorato and Esfeld (2010) for GRW.

In sum, the three mentioned primitive ontology theories of QM have been set out in the literature, but, as yet, research has not moved to assessing them in a comprehensive manner in order to find out which of them, if any, is most likely to be on the right track towards a fundamental ontology of the physical world. In the following, four guidelines for such an assessment are set out: (a) internal coherence; (b) empirical adequacy; (c) relationship to other theories; (d) explanatory value. Given the limits of a single paper, the assessment has to be brief and the reader has, whenever possible, to be referred to existing literature, without there being the space here to discuss that literature in detail.

## 2.    *Criterion I: Internal coherence*

Both BM and GRW have been criticized for being *ad hoc*, because they add something to textbook QM. As regards BM, this criticism concerns Bohm's formulation in the form of a second order theory, which adds a specific quantum force to the forces recognized in classical mechanics, known as the quantum potential (Bohm 1952, Bohm and Hiley 1993, Holland 1993). This quantum force can indeed be dismissed as *ad hoc*, because it does not match classical physics:

(i) Unlike a classical force, the quantum potential does not satisfy Newton's third law: there is no reaction from the particles that corresponds to the action of this force on them.

(ii) Unlike a classical force, the quantum potential does not necessarily decay as the spatial distance between the particles increases.

(iii) Unlike the classical forces of gravitation and electromagnetism, the quantum potential cannot be traced back to a property of the particles (like their mass or charge) whose effect is transmitted through a field. The quantum potential does not have a value at points in space-time, but only at points in configuration space.

(iv) Unlike in the whole of classical mechanics, in Bohm's theory, no second order equation fulfilling the scheme of Newton's second law (namely to employ forces that fix the temporal development of the velocity of particles) is needed in order to determine the motion of the particles. Instead, a first order equation in which the wave-function has the job to fix the velocity of the particles at a time $t$, given their position at $t$, is necessary and sufficient to obtain the Bohmian particle trajectories.

The – justified – criticism of the quantum potential does not hit today's dominant version of Bohm's theory, known as Bohmian mechanics (BM), which takes up de Broglie's (1928) original formulation of the theory with a law that is a first order equation.[5] There is no problem of internal coherence for this theory.

---

[4]   In the recent literature, Humeanism is also proposed as a stance that one can adopt with respect to the wave-function, notably in the context of BM (Miller 2013, Esfeld 2014, Callender 2014; Frigg and Hoefer 2007 argue for a Humean frequentist view of the GRW probabilities). Since there is no space to go into this issue here, it has to be left open here whether Humeanism also allows to consider the wave-function as representing a property of matter (as Miller 2013 tends to think) or whether on Humeanism, the significance of the wave-function is limited to offering an economical description of the distribution of matter in space-time (the primitive ontology) (as Esfeld 2014 assumes).

[5]   See Dürr, Goldstein and Zanghì (2013), and see Belousek (2003) and Solé (2013) for an overview of the different versions of Bohm's theory. See Bacciagaluppi and Valentini (2009, chs. 2 and 11) for a presentation and an assessment of de Broglie's original theory.

As regards the GRW theory, the criticism concerns the fact that this theory adds two new constants of nature to the Schrödinger equation. However, the introduction of these constants is well-motivated: the aim of the GRW theory is to improve on the collapse postulate in textbook QM, which is indeed *ad hoc*, since the notion of a measurement is not specified. The improvement consists in setting out one equation that matches by means of two additional constants what the collapse postulate seeks to capture, namely the fact of there being localized macrophysical objects. These constants concern the probability $\lambda$ with which the wave-function of a given quantum system undergoes a spontaneous localization (namely on average every $10^{15}$sec for a single elementary quantum system) and the width $\sigma$ of the localization (namely $10^{-7}$m). It is true that these new constants cannot be deduced from any fundamental principle, but have to be put in by taking the available empirical evidence as guideline. However, there is nothing wrong with figuring out the values of constants of nature on the basis of empirical evidence.

In sum, thus, none of the primitive ontology theories is hit by the objection of introducing *ad hoc* additions to textbook QM.

## 3.     Criterion II: Empirical adequacy

The theorems of Gleason (1957) and Kochen and Specker (1967) among others show that it is not possible to regard the quantum mechanical operators or observables as describing properties that the objects in nature possess, for one cannot attribute values to these observables independently of measurement contexts. That is one motivation for the primitive ontology theories: one cannot read an ontology of QM off from the textbook formalism, but has to put in a distribution of matter in physical space as the referent of that formalism. Against this background, regarding the position of something in physical space – be it particles, a matter density field, or flash-events – as the only local physical property is a consequent move. The formalism of the primitive ontology theories then sets out a law of the temporal development of the distribution of matter, with observables describing how the matter behaves in certain measurement contexts, that is, how the position of the objects in physical space develops in such contexts. Consider spin: these theories have to account for the outcomes of spin measurements in terms of the temporal development of the position of objects in physical space as described by the wave-function. This has been done in detail for BM (Bell 2004, ch. 4, Norsen 2013). There is no doubt that the same treatment is available for GRWm and GRWf (cf. Tumulka 2006, 2009). All these theories hence avoid the no-go theorems for additional, so-called hidden variables.

BM is empirically adequate as long as textbook QM is so, for it yields the same predictions for measurement outcomes as textbook QM, seeking to ground the textbook formalism in an ontology of particles and a law of motion for the particles. The GRW theory, by contrast, has to slightly deviate from the predictions of textbook QM, for it seeks to integrate what the collapse postulate stands for into a single equation by adding two constants to the Schrödinger equation (so that this equation is replaced with a non-linear and stochastic equation). By specifying an objective probability for spontaneous localization as well as a width of localization independently of measurements, a GRW-type equation cannot reproduce exactly what the collapse postulate achieves by a *fiat*. But these deviations are so slim that there currently are no experiments that test the GRW predictions against the ones of textbook QM.

Nonetheless, such experiments are in principle possible. If they were done and confirmed the predictions of the current GRW theory and contradicted the ones of textbook QM, that would create an entirely new situation. However, if they did not confirm the predictions of the current GRW theory, that would not automatically knock out this theory, since, as mentioned above, the constants added to the Schrödinger equation are figured out on the basis of the available empirical evidence. Hence, if new evidence becomes available, there is some room to adapt the constants that a GRW-type theory has to add to the Schrödinger equation, without infringing upon the falsifiability of such a theory (see Feldmann and Tumulka 2012). Consequently, it is not for sure that it will be possible to decide among the different primitive ontology theories of QM on the basis of experimental results.

There is another problem about the constants that GRW add to the Schrödinger equation raised in the literature, which concerns not only the empirical adequacy, but also the internal coherence of the GRW theory. The integration of what the collapse postulate stands for into the Schrödinger equation is achieved in such a way that after spontaneous localization, the wave-function takes the shape of a Gauss distribution, having a peak concentrated around a certain point in configuration space, but not being zero outside that peak. In the literature starting with Albert and Loewer (1996) and Lewis (1997), a concern is therefore expressed to the effect that the GRW theory does not achieve its aim, namely to describe measurement outcomes in the form of macrophysical objects having a definite position.

However, Tumulka (2011) convincingly argues that this objection puts the chart before the horse: the primitive ontology theories based on the GRW formalism *presuppose* an ontology of matter localized in physical space as the referent of the formalism, instead of seeking to derive such an ontology from the formalism. In other words, as already mentioned above, there are no superpositions of anything in the primitive ontology. Hence, instead of the spontaneous localization of the wave-function in configuration space being supposed to bring it about that matter is localized in physical space, the GRWm and the GRWf theory start from the assumption that there is an initial distribution of matter in physical space, and the GRW formalism only has the task to describe the temporal development of the initial distribution of matter localized in physical space (see already Monton 2004, pp. 418-419).

Nonetheless, there is a problem of empirical adequacy that hits GRWf. According to the GRWf ontology, macrophysical objects are, as Bell put it, galaxies of flash-events (Bell 2004, p. 205). Maudlin (2011, pp. 257-258) points out that this ontology implies that our standard conception of small classical objects such as DNA strands is radically false: if a DNA strand consists of about $10^9$ atoms, its wave-function undergoes a GRW hit roughly once a day, so that there is a configuration of flashes constituting a DNA strand only once a day, instead of there being a continuous object.

Against this background, consider the question of what a measuring apparatus interacts with when it is supposed to measure a quantum system with mass such as an electron. According to the solution to the measurement problem that GRW offers, the quantum system that is to be measured is coupled with the huge configuration of quantum systems that make up the measuring apparatus. It thereby becomes entangled with that huge configuration. That entanglement rapidly vanishes, since that huge configuration is immediately subject to a GRW hit. However, this account is not available for the flash ontology: there is nothing with which the measuring apparatus could interact or which could be coupled to it. There is no particle that enters it, and no field that gets in touch with it either. If the wave-function is a field, then, as mentioned above, it can only be a field on configuration space. Consequently, it cannot mediate the interaction among the flashes in space-time (as a classical field could do).

If one attributes to an initial configuration of flashes – to take a simple example, consider the two flashes at the source of the EPR experiment – the propensity to bring about further flashes, this propensity may manifest itself spontaneously in the production of new flashes. However, as regards the EPR experiment, one may have to wait billions of years for such a spontaneous manifestation to occur. If one wants to trigger that propensity by setting up a galaxy of flashes that forms a measuring apparatus, the problem is that the GRWf ontology does not have the means at its disposal to give a physical answer to the question of how the occurrence of new flashes could be triggered by placing galaxies of flashes in the form of measuring apparatuses at certain locations in space and time: there is at almost all spatio-temporal locations nothing with which such an apparatus could interact in order to trigger the occurrence of new flashes that could make up a measurement outcome. In other words, there is almost nowhere in physical space-time (by contrast to configuration space) anything that could distinguish a particular place and time to set up a measuring apparatus so that the occurrence of new flashes, constituting a measurement outcome, would be triggered.

Hence, even before it comes to a physical account of interactions (which is as yet not worked out in GRWf), the theory faces the objection that the flash ontology is too sparse: it does not have the means at its disposal to account for interactions that are supposed to trigger the occurrence of further flashes. In sum, thus, there is a serious problem for the GRWf theory, but both BM and GRWm fare well as far as empirical adequacy is concerned.

## *4.*     *Criterion III: Relationship to other theories*

The primitive ontology theories are well worked out for non-relativistic QM. The main concern is how these theories can accommodate relativistic space-time. The special theory of relativity implements Lorentz-invariance in the following sense: there is no globally preferred reference frame and thus no privileged foliation of four-dimensional space-time into three-dimensional spatial hypersurfaces that are ordered in one-dimensional time. Hence, there is no absolute simultaneity. However, it seems that BM and GRW are committed to absolute simultaneity: in BM, the velocity of any particle at a time $t$ depends, strictly speaking, on the position of all the other particles at that very $t$. In GRW, the spontaneous localization of the wave-function in configuration space describes a change in the position of the whole distribution of matter in space at once.

The question of the compatibility of QM and special relativity concerns not only the primitive ontology theories, it is a general one. Bell's theorem (1964) (reprinted in Bell 2004, ch. 2) establishes that any theory that is to match the experimentally confirmed predictions of textbook QM has to violate a principle of locality: in certain situations – such as the EPR experiment –, the probabilities for what happens at a given space-time point are not entirely fixed by what there is in the past light-cone of that point, but depend also on what measurements take place at a spacelike separated interval (see Bell 2004, ch. 24, and Seevinck and Uffink 2011 for precisions). Bell's theorem is therefore widely taken to show that events that are separated from a given space-time point by a spacelike interval can contribute to determining what happens at that point. Maudlin (2011, chs. 1-6) analyses this determination in terms of non-local influences (see in particular pp. 118-119, 135-141).

The most significant progress in bringing primitive ontology theories of QM together with special relativity has been achieved by Tumulka (2006, 2009). He has set out a relativistic version of the GRWf ontology (rGRWf): he considers an initial configuration of flashes and an initial wave-function attributed to that configuration (formulated with respect to an arbitrary hypersurface). He puts that wave-function into the GRW equation (more precisely the GRW amendment of the Dirac equation), obtaining entire distributions of flashes in space-time with a certain probability attached to each such distribution. He shows that any such distribution as a whole can be represented in a Lorentz-invariant manner, because the rGRWf law is Lorentz-invariant. Furthermore, the rGRWf theory makes it possible to attribute a unique wave-function to the configuration of flashes on each particular spatial hypersurface given that hypersurface and all the flashes that occur earlier than it, and given the wave-function specified relative to any earlier hypersurface. The transition from the wave-function of the configuration of flashes on one hypersurface to the wave-function of the configuration of flashes on any later hypersurface also is Lorentz-invariant in the sense that it can be defined independently of the particular way in which the intervening space-time is foliated. Recent research seeks to extend these results from the GRWf ontology to the GRWm ontology (see notably Bedingham et al. 2013).

However, Esfeld and Gisin (2014) have raised the following objection: it is not possible to conceive the coming into being of the individual flashes in a Lorentz-invariant manner. The reason is that the occurrence of some flashes depends on where in space-time other flashes occur: considering the EPR experiment with Alice and Bob carrying out spin measurements in the two wings of the experiment, in one frame, Alice's outcome flash is independent of the flashes that constitute Bob's setting and outcome; in another frame, Alice's outcome flash depends on the flashes that constitute Bob's setting and outcome. The same goes for Bob's outcome flash. Esfeld and Gisin (2014) therefore argue on the basis of Bell's theorem that there is no Lorentz-invariant theory of these dependency relations possible.

This argument generalizes from GRWf to all primitive ontology theories: as long as one considers *only* the primitive ontology – that is, takes the ontology to consist *only* in the distribution of matter (particles, matter density, flashes) throughout the *whole* of space-time –, there is no problem with Lorentz-invariance. However, as soon as one takes the entanglement of the wave-function in configuration space to refer to dependency relations among some elements of the primitive ontology in space-time, Bell's theorem implies that there is no Lorentz-invariant theory of these dependency relations possible, since there then is a fact for any given flash, particle position or matter density value at a space-time point of whether or not the occurrence of that flash, particle position or matter density value depends on where flashes, particle positions or matter density values occur at spacelike separated locations.

Against this background, on the one hand, it is still a live option to maintain that the standard concern about the primitive ontology theories, namely that they are committed to more space-time structure than recognized by the special theory of relativity, is well-grounded, despite the results achieved by Tumulka (2006, 2009) and Bedingham et al. (2013). On the other hand, even if one upholds this concern, it does not constitute an objection against the primitive ontology theories in particular: to the extent that there is a commitment to additional space-time structure, this is a consequence of adopting realism with respect to the dependency relations between space-like separated events brought out by Bell's theorem. The decisive open issue therefore is whether these theories can introduce such additional space-time structure in a non *ad hoc* manner. Thus, Dürr et al. (2013) suggest in the context of BM that the additional space-time structure follows from the universal wave-function.

If one takes quantum field theory (QFT) into consideration, nothing of substance changes as far as this assessment is concerned. Bell's theorem applies to QFT in the same manner as to non-relativistic QM.[6] That is to say, QFT by no means solves the problem of how to accommodate the non-local quantum correlations in a relativistic space-time. Although the primitive ontology theories are worked out in the context of QM, there is no principled problem in extending them to QFT. Thus, Tumulka's rGRWf theory is already formulated by employing a GRW-version of the Dirac equation, and it seems that the same is possible in the framework of the GRWm theory. As regards BM, there are currently several proposals for a Bohmian QFT (see Struyve 2011 for an overview), with the one going back to Bell (2004, ch. 19) being the most well worked out one, producing the same predictions as textbook QFT (see Dürr, Goldstein and Zanghì 2013, ch. 10).

The big open issue is how the primitive ontology theories fare when it comes to the relationship between quantum theory and general relativity, that is, the search for a theory of quantum gravity (QG). If it should prove sound that in QG, the Wheeler-deWitt equation takes the place of the Schrödinger equation in QM, this would knock out GRW-type theories as candidates for providing the dynamics of fundamental physical systems, because in the Wheeler-deWitt equation figures a stationary, universal wave-function. Hence, that equation cannot be amended by adding constants to it that concern the temporal development of the wave-function towards spontaneous localization, for there is no development of the wave-function. In this case, both GRWm and GRWf would be ruled out.

BM, by contrast, can accommodate a universal wave-function that is stationary, since the temporal development of the primitive ontology is not given by the Schrödinger equation, but by another equation. The Bohmian guiding equation can contain a stationary wave-function and yet describe the – temporal – development of a primitive ontology. The challenge for the Bohmian hence is to come up with a proposal for a primitive ontology and a guiding equation in the domain of QG.[7]

In any case, since there is no established physical theory of QG as yet, one cannot make the assessment of the primitive ontology theories of quantum physics dependent on how they would fare in this domain – and it would be unreasonable to suspend their assessment until the day physics may have reached a final theory of matter and space-time. In sum, thus, there are open issues in the relationship between primitive ontology theories of QM and relativity physics, but as things stand no cogent objection from relativity physics that hits these theories in particular. Furthermore, there is nothing in this domain that constitutes a decisive advantage of one of these theories (such as GRWf) over the other ones.

## 5.    *Criterion IV: Explanatory value*

A central motivation for the primitive ontology theories is that by accommodating the existence of measurement outcomes in a quantum formalism, these theories offer the prospect of explaining these outcomes, instead of simply providing a tool for calculating probabilities for them. Explaining them does not necessarily mean giving a reason why a particular outcome is obtained in a particular experiment. It means answering the question of why the textbook QM probabilities hold.

---

[6]   See Bell (2004, ch. 24) and Hofer-Szabó and Vescernyés (2013) as well as Lazarovici (2013) for the current discussion.

[7]   For first sketches in that direction, see notably Goldstein and Teufel (2001, reprinted in Dürr, Goldstein and Zanghì 2013, ch. 11) as well as Vassallo and Esfeld (2014).

A GRW-type theory answers this question by putting forward a probabilistic law in which new constants of nature figure, with these constants explaining the QM probabilities. Thus, on the view of laws being grounded in dispositions as applied to GRW in Dorato and Esfeld (2010), the matter density has a propensity for spontaneous localization (GRWm) and a configuration of flashes has the propensity to bring about further flashes (GRWf). These propensities ground the GRW probabilities and thereby the formalism of textbook QM.[8]

BM, by contrast, is a deterministic theory. Both the Schrödinger equation and the guiding equation are deterministic laws. BM, however, is not wedded to determinism: in what is known as Bell-type QFT, there is a stochastic guiding equation (see Dürr, Goldstein and Zanghì 2013, ch. 10). Nonetheless, as far as the explanation of the QM textbook formalism is concerned, BM can introduce probabilities only in the way in which probabilities enter into classical statistical mechanics, namely by referring to ignorance of the exact initial conditions. This means that probabilities are not grounded in the ontology of BM (the primitive ontology and the law of motion), but that a link from that ontology to probabilities has to be established. In any primitive ontology theory of QM, the wave-function has a double role, namely (a) to figure in the law that describes the temporal development of the distribution of matter in physical space and (b) to enable the calculation of probabilities for measurement outcomes. In a theory with a stochastic law, both these roles can be incorporated in a single equation. In a theory with a deterministic law for (a), further argumentation is needed to capture (b).

On the basis of assuming that the initial particle configuration of the universe is typical in a precise mathematical sense, Dürr, Goldstein and Zanghì (2013, ch. 2, originally published 1992) introduce what is known as the quantum equilibrium hypothesis, which then enables the deduction of the QM probabilities in BM. Despite the criticism raised notably by Bricmont (2001), Callender (2007) gives cogent reasons for regarding this deduction as successful. A strong case can hence be made for both BM on the one hand and GRWm and GRWf on the other being in the position to explain why the QM probabilities obtain.

That notwithstanding, the explanation that is expected from primitive ontology theories is not limited to answering the question of why the QM probabilities hold. A more fundamental explanation is called for, namely one that makes the most astonishing feature of QM intelligible, that is, the non-locality brought out by the experiments following Bell's theorem. Explaining quantum non-locality, however, does not mean answering the question of why nature is non-local. Non-locality, if it obtains, is a fundamental feature of nature. Thus, there can be no explanation of why nature is non-local, in the same way as there can be no explanation of why nature is local, if it were local (as was assumed in classical field theories). What can with reason be called for is an account of non-locality in the sense of providing an ontology that accommodates the empirical evidence for quantum non-locality.

Einstein has set up the standard that such an account has to meet. For Einstein, non-locality amounted to "spooky action at a distance". Consider a simple example that illustrates what Einstein meant by this remark, namely the thought experiment with one particle in a box that

---

8    See Suárez (2007) for the propensity theory of probabilities applied to QM in general.

Einstein presented at the Solvay conference in 1927 (my presentation is based on de Broglie's version of the thought experiment in de Broglie 1964, pp. 28-29, and on Norsen 2005): the box is split in two halves which are sent in opposite directions, say from Brussels to Paris and Tokyo. When the half-box arriving in Tokyo is opened and found to be empty, there is on all accounts of QM that acknowledge that measurements have outcomes a fact that the particle is in the half-box in Paris. If one takes the textbooks' postulate of the collapse of the wave-function upon measurement literally, this means that the act of opening the box in Tokyo creates the fact that there is a particle in the box in Paris. This is an illustration of what Einstein regarded as "spooky action at a distance". Note that this is not like the Newtonian force of gravitation, which spreads instantaneously all over space, but decreases the larger the spatial distance from the source is. In this case, a local operation in one place creates a particular fact in exactly one other place, however far apart in space that other place may be. Since such action at a distance is absurd according to Einstein, he maintained that the particle always travels in one of the two half-boxes, depending on its initial position, and that its motion is not influenced by whatever operation is carried out on the other half-box.

In this case, such a local account is indeed available and provided by BM: according to BM, there always is one particle moving on a continuous trajectory in one of the two half-boxes, and opening one of them only reveals where the particle was all the time. On GRWm, by contrast, the particle is in fact a matter density field that stretches over the whole box and that is split in two halves of equal density when the box is split, these matter densities travelling in opposite directions. Upon interaction with a measurement device, one of these matter densities (the one in Tokyo in the example given above) vanishes, while the matter density in the other half-box (the one in Paris) increases so that the whole matter is concentrated in one of the half-boxes. One might be tempted to say that some matter travels from Tokyo to Paris, but since it is impossible to assign any finite speed to this travel, the use of the term "travel" is inappropriate. For lack of a better expression let us say that some matter is delocated from Tokyo to Paris; for even if the spontaneous localization of the wave-function in configuration space is a continuous process, the time it takes for the matter density to disappear in one place and to reappear in another place does not depend on the distance between the two places. On GRWf, there only is one flash in the box at the source of the experiment in Brussels, nothing travels in either box, but opening the box in Tokyo triggers the occurrence of a new flash in Paris.

When moving from Einstein's thought experiment with one particle in a box (1927) to the EPR experiment (Einstein, Podolsky and Rosen 1935), nothing of substance changes in the two GRW ontologies. On GRWm, again, the measurement in one wing of the experiment triggers a delocation of the matter density, more precisely a change in its shape in both wings of the experiment, so that, in Bohm's (1951) version of the experiment, the shape of the matter density constitutes two spin measurement outcomes. On GRWf, again, there are only two flashes at the source and nothing travelling in either wing of the experiment, but the measurement in one wing triggers the occurrence of two new flashes, one in each wing of the experiment, such that the two flashes make up for two spin measurement outcomes.

As regards BM, whereas BM gives a local account of the thought experiment with the particle in a box, it cannot do so in the EPR experiment. On BM, fixing the parameter in one wing of this experiment influences the trajectory of the particles in both wings (BM being a deterministic theory, it has to violate the condition known as parameter independence, since on a deterministic theory, it is determined what the outcome of a measurement is before the measurement actually occurs). But there is no direct interaction among the particles on BM. Fixing the parameter in one wing of the EPR experiment influences the trajectory of the particles in both wings via the wave-function of the whole system, which has the job to determine the velocity of each particle at *t*, given the position of all the particles at *t*. Nonetheless, BM supports counterfactuals of the type: "If Alice had chosen a different setting, Bob would have obtained a different outcome". But there is no direct influence from Alice's setting to Bob's outcome. Alice's setting influences Bob's outcome only via the wave-function, which in turn influences *both* outcomes. Thus, strictly speaking, on BM, the velocity of any particle at *t* depends on the position of *all* the other particles at *t* via the wave-function (including the particles that constitute the settings of the parameters and the measurement apparatuses). However, due to the decoherence of the universal wave-function, in many situations, the position of distant particles is *de facto* irrelevant for the trajectory of a given particle and what is known as an effective wave-function can be attributed to the particle (as in the case of the particle in a box).

Consequently, if one conceives the universal wave-function in BM as representing a property of the particles that determines their velocity (see end of section 1 above), BM is able to provide an ontology that makes quantum non-locality intelligible, thereby rebutting Einstein's view that a failure of locality would imply "spooky action at a distance": there is a holistic property of all the particles taken together that fixes their velocity. There is no action at a distance among anything, simply because a holistic property instantiated by the configuration of matter is another conception of the determination of the movement of something than direct interaction among the parts of that configuration. Hence, there is quantum non-locality because the temporal development of quantum systems is not determined by properties that are intrinsic to each object (as mass and charge are intrinsic properties of particles in classical mechanics), but by a holistic property that is instantiated by the configuration of all the quantum objects. This explanation may sound unfamiliar given our familiarity with a classical domain of objects being moved by local forces, but it is complete and coherent. As already pointed out by Solé (2013) in another context, there is no need for a formulation in terms of a second order theory that is committed to a dubious quantum force (see section 2 above) for BM to be explanatory.

As mentioned above, there can be no explanation of why nature is non-local, if it is non-local, as there can be no explanation of why nature is local, if it is local. In the latter case, there are local forces going back to intrinsic properties of the objects in nature; in the former case, there is a holistic property instantiated by the configuration of the objects. As there can be no further explanation of how intrinsic properties of particles such as their mass and charge give rise to forces such as gravitation and electromagnetism that accelerate the particles, so there can be no further explanation of how a holistic property instantiated by the configuration of the particles determines the velocity of the particles.

Coming back to GRWm and GRWf, one can also in these theories conceive the wave-function as representing a holistic property, namely a propensity of the matter density or a configuration of flashes for spontaneous contraction or the production of new flashes (see end of section 1 above). But this on its own is not sufficient as an account of quantum non-locality. The crucial point is how the actualization of this propensity is conceived. In GRWf, as the example of the particle in a box illustrates, a local operation in one region of space can actualize this propensity by producing the occurrence of one single flash at a spacelike separated location. By way of consequence, it is difficult to see how GRWf could avoid the conclusion of this actualization amounting to "spooky action at a distance": in the case of Einstein's boxes, a local operation that is not an interaction with any quantum system (since there simply is no quantum system) brings about exactly one quantum flash not at the location where this interaction takes place, but at a location that can be arbitrarily far apart in space. This problem is a consequence of the one elaborated on at the end of section 3, namely that the GRWf ontology has a difficulty in making intelligible how new flashes can be triggered through interventions such as measuring operations (by contrast to new flashes simply occurring spontaneously if one is prepared to wait long enough).

Against the background of the discussion of the relationship between quantum non-locality and relativity physics in the preceding section, we can make the following point in this context: in BM and GRWm, there is *spatial* non-locality in that what occurs at a given point in space depends on what occurs at other points in space *at the same time* (while these other points can be arbitrarily far apart in space); in GRWf, there also is *temporal* non-locality in that the propensity for new flashes to occur and its actualization encompasses not only distances in space that can be arbitrarily large, but also distances in time that can be arbitrarily big (in the extreme case, there may simply be no flashes at all at a given time, so nothing that could instantiate the propensity for new flashes to occur). This feature of GRWf is problematic for the following reason: as sketched out above, there is an account available that makes spatial non-locality intelligible in terms of a holistic property that is instantiated by the configuration of matter as a whole at any given time (by contrast to local, intrinsic properties of parts of the configuration of matter like the mass and charge of particles in classical mechanics). However, it is not clear whether and how this account could be extended to cover temporal non-locality as well. At least, countenancing a fundamental physical property whose instantiation and actualization stretches over gaps in time seems to be much more problematic than endorsing a property that is instantiated at a time (albeit by the whole configuration of matter at that time).

Turning to the features that distinguish GRWm from BM in the account of spatial non-locality, over and above of what is admitted in BM, according to GRWm, the mentioned propensity is actualized in the form of a delocation of parts of the matter density. This delocation introduces a feature into the account of quantum non-locality that is absent in BM and that is problematic for the following reason: as the example of the particle in a box shows, one and the same matter density can disappear in Tokyo and reappear in Paris without travelling from Tokyo to Paris. This is not action at a distance, since the action is local in Tokyo, but the process of the delocation of matter without anything travelling from one place to another place can with good reason be considered as mysterious.

In sum, the three proposals for a primitive ontology of quantum physics contain three different accounts of non-locality:

*Particles*: they are always there and always move on continuous trajectories that are determined by a holistic property instantiated by the particle configuration at any time.

*Matter density*: it is always there all over space. Its temporal development is determined by the actualization of a holistic propensity instantiated by the matter density at any time. However, this actualization implies that one and the same matter disappears in location *A* and reappears in location *B* without travelling from *A* to *B*.

*Flashes*: there is a holistic propensity of configurations of flashes to produce new flashes. However, in many situations, there are no flashes there that could instantiate such a propensity. This theory therefore is committed to the view that this propensity is instantiated and actualized by flashes over distances in space as well as in time that can be arbitrarily big.

In other words, it seems that theories whose dynamics is non-local also in configuration space (as is the GRW dynamics with the spontaneous localization of the wave-function) cannot avoid some sort of the pitfall of what Einstein termed "spooky action at a distance": on such theories, the non-local dynamics in configuration space either signifies direct action at a distance (GRWf) or the delocation of some matter all over space (GRWm). By contrast, a theory with a local dynamics in configuration space (development of the wave-function according to the Schrödinger equation) has the conceptual means at its disposal to avoid the commitment to "spooky action at a distance" and yet recognize the existence of measurement outcomes: such a theory can conceive objects that always move on continuous trajectories (without anything ever being delocated across space), with these trajectories being determined by a holistic property instantiated by the configuration of these objects at any time.

## 6. *Conclusion*

This paper has made a case for the following conclusions: the standard objections against the primitive ontology theories of QM do not hold up to scrutiny. These theories pass the tests of internal coherence and empirical adequacy, and they do not face a problem stemming from relativity physics for which they are responsible. Furthermore, they can with reason claim to offer an explanation of the QM textbook formalism for calculating probabilities of measurement outcomes. However, the GRWf theory has a serious problem in making interactions that trigger the occurrence of new flashes intelligible. In general, the success of these theories depends upon whether or not they can achieve an account that makes quantum non-locality intelligible, avoiding the features of what Einstein dismissed as "spooky action at a distance". BM can do so, whereas as things stand it is difficult to see how GRWf and GRWm could do so. Hence, if one acknowledges that there is a good reason to go for a primitive ontology theory of quantum physics, BM is at the current state of the art the best candidate for a theory that may be on the right track towards a fundamental ontology of the physical world, whereas GRWm and GRWf face serious difficulties in achieving an account of quantum non-locality that can stand firm.